\documentclass{article}

\usepackage{microtype}
\usepackage{graphicx}
\usepackage{subfigure}
\usepackage{booktabs} % for professional tables
\usepackage{amsmath}
\usepackage{amsfonts}
\usepackage{enumitem}

\usepackage{hyperref}

\usepackage[accepted]{icml2020}

\newcommand{\minihead}[1]{{\vspace{.5em}\noindent\textbf{#1.} }}

\icmltitlerunning{Scaling up Trustless DNN Inference with Zero-Knowledge
Proofs}

\begin{document}

\twocolumn[
\icmltitle{Scaling up Trustless DNN Inference with  Zero-Knowledge
Proofs}

\icmlsetsymbol{equal}{*}

\begin{icmlauthorlist}
\icmlauthor{Daniel Kang}{uiuc}
\icmlauthor{Tatsunori Hashimoto}{stanford}
\icmlauthor{Ion Stoica}{berkeley}
\icmlauthor{Yi Sun}{chicago}
\end{icmlauthorlist}

\icmlaffiliation{uiuc}{University of Illinois, Urbana-Champaign}
\icmlaffiliation{stanford}{Stanford University}
\icmlaffiliation{chicago}{University of Chicago}
\icmlaffiliation{berkeley}{University of California, Berkeley}

\icmlcorrespondingauthor{Daniel Kang}{ddkang@illinois.edu}

\vskip 0.3in
]

\printAffiliationsAndNotice{} % otherwise use the standard text.

\begin{abstract}

As ML models have increased in capabilities and accuracy, so has the complexity
of their deployments. Increasingly, ML model consumers are turning to service
providers to serve the ML models in the ML-as-a-service (MLaaS) paradigm. As
MLaaS proliferates, a critical requirement emerges: how can model consumers
verify that the correct predictions were served, in the face of malicious, lazy,
or buggy service providers?

In this work, we present the first ImageNet-scale method to verify ML model
execution non-interactively. To do so, we leverage recent developments in
ZK-SNARKs (zero-knowledge succinct non-interactive argument of knowledge), a
form of zero-knowledge proofs. ZK-SNARKs allows us to verify ML model execution
non-interactively and with \emph{only} standard cryptographic hardness
assumptions. In particular, we provide the first ZK-SNARK proof of valid
inference for a full resolution ImageNet model, achieving 79\% top-5 accuracy. 
We further use these ZK-SNARKs to design protocols to verify ML model execution
in a variety of scenarios, including for verifying MLaaS predictions, verifying
MLaaS model accuracy, and using ML models for trustless retrieval. Together, our
results show that ZK-SNARKs have the promise to make verified ML model inference
practical.

\end{abstract}

\section{Introduction}

ML models have been increasing in capability and accuracy. In tandem, the
complexity of ML deployments has also been exploding. As a result, many
consumers of ML models now outsource the training and inference of ML models to
service providers, which is typically called ``ML-as-a-service'' (MLaaS). MLaaS
providers are proliferating, from major cloud vendors (e.g., Amazon, Google,
Microsoft, OpenAI) to startups (e.g., NLPCloud, BigML).

A critical requirement emerges as MLaaS providers become more prevalent: how can
the model consumer (MC) verify that the model provider (MP) has correctly served
predictions? In particular, these MPs execute model inference in untrusted
environments from the perspective of the MC. In the untrusted setting, these MPs
may be lazy (i.e., serve random predictions), dishonest (i.e., serve malicious
predictions), or inadvertently serve incorrect predictions (e.g., through bugs
in serving code).

% As a parallel trend, there are increasingly many circumstances where ML model
% inference must be verified on hidden inputs (although the model itself may be
% shared). For example, consider the setting where a journalist issues a Freedom
% of Information Act (FOIA) request where the journalist desires to find documents
% or images that match a semantic pattern. For example, the requester may be
% interested in finding images that contain hate symbols. The requester has an ML
% model they wish to run on the government data to select these images. However,
% the government wishes to keep the images hidden (except the requested images).
% How can the requester execute the ML model on the data while the government
% keeps the images hidden?

In this work, we propose using the cryptographic primitive of ZK-SNARKs
(Zero-Knowledge Succinct Non-Interactive Argument of Knowledge) to address the
problem of practically verifying ML model execution in untrusted settings. We
present the first ZK-SNARK circuits that can verify inference for ImageNet-scale
models, in contrast to prior work that is limited to toy datasets such as MNIST
or CIFAR-10 \cite{feng2021zen, weng2022pvcnn, lee2020vcnn, liu2021zkcnn}. We are
able to verify a proof of valid inference for MobileNet v2 achieving 79\%
accuracy while simultaneously being verifiable in 10 seconds on commodity
hardware. Furthermore, our proving times can improve up to one to four orders of
magnitude compared to prior work \cite{feng2021zen, weng2022pvcnn, lee2020vcnn,
liu2021zkcnn}. We further provide practical protocols leveraging these ZK-SNARKs
to verify ML model accuracy, verify MP predictions, and using ML models for
audits. These results demonstrate the feasibility of practical, verified ML
model execution.

ZK-SNARKs are a cryptographic primitive in which a party can provide a
certificate of the execution of a computation such that no information about the
inputs or intermediate steps of the computation are revealed to other parties.
ZK-SNARKs have a number of surprising properties (Section~\ref{sec:snarks}).
Importantly for verified DNN execution, ZK-SNARKs allow portions of the input
and intermediates to be kept hidden (while selectively revealing certain inputs)
and are \emph{non-interactive}. The non-interactivity allows third parties to
trustlessly adjudicate disputes between MPs and MCs and verify the computation
without participating in the computation itself.

In the setting of verified DNN inference, the weights, inputs, or neither can be
made public while keeping the others hidden. The hidden portions can then be
committed to by computing and revealing hashes of the inputs, weights, or both
(respectively). In particular, a MP may be interested in keeping its proprietary
weights hidden while being able to convince a MC of valid inference. The
ZK-SNARK primitive allows the MP to commit to the (hidden) weights while proving
execution.

To ZK-SNARK ImageNet-scale models, we leverage recent developments in ZK-SNARK
proving systems \cite{halo2}. Our key insight is that \emph{off-the-shelf}
proving systems for generic computation are sufficient for verified ML model
execution, with careful arithmetization (i.e., translation) from DNN
specifications to ZK-SNARK arithmetic circuits. Our arithmetization uses two
novel optimizations: lookup arguments for non-linearities and reuse of
sub-circuits across layers (Section~\ref{sec:opts}). Without our optimizations,
the ZK-SNARK construction will require an impractically large amount of hardware
resources.

% In order to ZK-SNARK ImageNet-scale models, we build a translation layer from
% DNN specifications to ZK-SNARK arithmetic circuits. Furthermore, to efficiently
% ZK-SNARK models, we propose novel methods of using new arithmetization methods.
% To do so, we efficiently use lookup arguments and share resources between DNN
% layers. Without these optimizations, the ZK-SNARKs construction results in
% out-of-memory errors on our hardware platform.

Given the ability to ZK-SNARK ML models while committing to and selectively
revealing chosen portions of their inputs, we propose methods of verifying MLaaS
model accuracy, MLaaS model predictions, and trustless retrieval of documents in
the face of \emph{malicious} adversaries. Our protocols combine ZK-SNARK proofs
and economic incentives to create trustless systems for these tasks. We further
provide cost estimates for executing these protocols.

% These protocols work by forcing the MP or MC to commit to the input
% data and/or weights and verifying the results of the computation. However, each
% protocol requires a different set of commitments and economic incentives to
% operate.

In summary, our contributions are:
\begin{enumerate}[topsep=0.1em, leftmargin=1.3em]
  \item The first ImageNet-scale ZK-SNARK circuit that can be proved and
  verified on commodity hardware (Section~\ref{sec:eval}).
  \item Novel arithmetization optimizations for DNN inference in the form of
  lookup arguments for non-linearities and sub-circuit reuse to enable
  ImageNet-scale ZK-SNARKs (Section~\ref{sec:opts}).
  \item Protocols and proofs of concept for leveraging these ZK-SNARKs in
  methods for auditing via ML models, verifying ML model accuracy, and serving
  ML model predictions in the face of adversaries (Section~\ref{sec:protocols}).
\end{enumerate}

% In the remainder of the manuscript, we describe ZK-SNARKs in further detail,
% our protocols that leverage ZK-SNARKS, and our ZK-SNARK translation layer and
% its optimizations.

\section{Related Work}

\minihead{Secure ML}
Recent work has proposed secure ML as a paradigm for executing ML models
\cite{ghodsi2017safetynets, mohassel2017secureml, knott2021crypten}. There are a
wide range of security models, including verifying execution of a known model on
untrusted clouds \cite{ghodsi2017safetynets}, input privacy-preserving inference
\cite{knott2021crypten}, and weight privacy-preserving inference. The most
common methods of doing secure ML are with multi-party computation (MPC),
homomorphic encryption (HE), or interactive proofs (IPs). As we describe, these
methods are either impractical, do not work in the face of malicious adversaries
\cite{knott2021crypten, kumar2020cryptflow, lam2022tabula, mishra2020delphi}, or
do not hide the weights/inputs \cite{ghodsi2017safetynets}. In this work, we
propose practical methods of doing verified ML execution in the face of
malicious adversaries.

\minihead{MPC}
One of the most common methods of doing secure ML is with MPCs, in which the
computation is shared across multiple parties \cite{knott2021crypten,
kumar2020cryptflow, lam2022tabula, mishra2020delphi, jha2021deepreduce}. There
are a variety of MPC protocols with different guarantees. However, all MPC
protocols have shared properties: they require interaction (i.e., both parties
must be simultaneously online) but can perform computation without revealing the
computation inputs (i.e., weights and ML model inputs) across parties.

There are several security assumptions for different MPC protocols. The most
common security assumption is the \emph{semi-honest adversary}, in which the
malicious party participates in the protocol honestly but attempts to steal
information.  In this work, we focus on potentially \emph{malicious
adversaries}, who can choose to deviate from the protocol. Unfortunately, MPC
that is secure against malicious adversaries is impractical: it can cost up to
550 GB of communication and 657 seconds of compute per example on toy datasets
\cite{pentyala2021privacy}. In this work, we provide a practical, alternative
method of verifying ML model inference in the face of malicious adversaries.
Furthermore, our methods do not require per-example communication.

\minihead{HE}
Homomorphic encryption allows parties to perform computations on encrypted data
without first decrypting the data \cite{armknecht2015guide}. HE is deployed to
preserve privacy of the inputs, but cannot be used to verify that ML model
execution happened correctly. Furthermore, HE is incredibly expensive. Since ML
model inference can take up to gigaflops of computation, HE for ML model
inference is currently impractical, only working on toy datasets such as MNIST
or CIFAR-10 \cite{lou2021hemet, juvekar2018gazelle}.

\minihead{ZK-SNARKs for secure ML}
Some recent work has produced ZK-SNARK protocols for neural network inference on
smaller datasets like MNIST and CIFAR-10. Some of these works like \cite{feng2021zen}
use older proving systems like \cite{groth2016size}. Other works
\cite{ghodsi2017safety, lee2020vcnn, liu2021zkcnn, weng2022pvcnn} use interactive
proof or ZK-SNARK protocols based on sum-check \cite{thaler2013time} custom-tailored
to DNN operations such as convolutions or matrix multiplications. Compared to these
works, our work in the modern Halo2 proving system \cite{halo2} allows us to use
the Plonkish arithmetization to more efficiently represent DNN inference by leveraging
lookup arguments and well-defined custom gates.  Combined with the efficient
software package \texttt{halo2} and advances in automatic translation, we are
able to outperform these methods.

\section{ZK-SNARKs}
\label{sec:snarks}

\minihead{Overview}
Consider the task of verifying a function evaluation $y = f(x; w)$ with
\emph{public inputs} $x$, \emph{private inputs} $w$, and output $y$. For
example, in the setting of public input and hidden model, $x$ may be an image,
$w$ may be the weights of a DNN, and $y$ may be the result of executing the DNN
with weights $w$ on $x$.

A ZK-SNARK \cite{bitansky2017hunting} is a cryptographic protocol allowing a
Prover to generate a proof $\pi$ so that with knowledge of $\pi$, $y$, and $x$
alone, a Verifier can check that the Prover knows some $w$ so that $y = f(x;
w)$. ZK-SNARK protocols satisfy several non-intuitive properties summarized
informally below:
\begin{enumerate}
  \item \emph{Succinctness}: The proof size is sub-linear (typically constant or
  logarithmic) in the size of the computation (i.e., complexity of $f$).
  \item \emph{Non-interactivity}: Proof generation does not require interaction
  between the verifier and prover.
  \item \emph{Knowledge soundess}: A computationally bounded prover cannot
  generate proofs for incorrect executions.
  \item \emph{Completeness}: Proofs of correct execution verify successfully.
  \item \emph{Zero-knowledge}: The proof reveals no information about
  private inputs beyond what is contained in the output and public inputs.
\end{enumerate}

Most ZK-SNARK protocols proceed in two steps. In the first step, called
\emph{arithmetization}, they produce a system of polynomial equations over a
large prime field (an arithmetic circuit) so that finding a solution is
equivalent to computing $f(x; w)$. Namely, for $(f, y, x, w)$, the circuit
constraints are met if and only if $y = f(x; w)$. In the second step, a
cryptographic \emph{proof system}, often called a \emph{backend}, is used
to generate a ZK-SNARK proof. 

This work uses the Halo2 ZK-SNARK protocol \cite{halo2} implemented in the
\texttt{halo2} software package.  In contrast to ZK-SNARK schemes custom
designed for neural networks in prior work \cite{liu2021zkcnn, lee2020vcnn},
Halo2 is designed for general-purpose computation, and \texttt{halo2} has a
broader developer ecosystem.  This means we inherit the security,
efficiency, and usability of the resulting developer tooling.  In the remainder
of this section, we describe the arithmetization and other properties of Halo2.

\minihead{Plonkish arithmetization}
Halo2 uses the Plonkish arithmetization \cite{halo2}, which allows polynomial
constraints with certain restricted forms of randomness.  It is a special case
of a randomized AIR with preprocessing \cite{bensasson2018scalable, gabizon2021from}
which unifies unifies recent proof systems including PlonK, plookup, and PlonKup
 \cite{gabizon2019plonk, gabizon2020plookup, pearson2022plonkup}.

% a subset of randomized
% AIR with preprocessing \cite{bensasson2018scalable, gabizon2021from} unifying
% the recent proof systems PlonK, plookup, and PlonKup
% \cite{gabizon2019plonk, gabizon2020plookup, pearson2022plonkup}.

Variables in the arithmetic circuit are arranged in a rectangular grid with
cells valued in a $254$-bit prime field. The Plonkish arithmetization allows
three types of constraints with which any computation may be expressed:\footnote{Any
arbitrary computation can be expressed, but the size of the arithmetized circuit
depends heavily on the nature of the computation.}

\emph{Custom gates} are polynomial expressions over cells in a single row which
must vanish on \emph{all} rows of the grid. As a simple example, consider a grid
with columns labeled $a, b, c$ with $a_i, b_i, c_i$ being the cells in row $i$.
The custom \emph{multiplication gate}
\[
a_i \cdot b_i - c_i = 0
\]
enforces that $c_i = a_i \cdot b_i$ for all rows $i$.

In nearly all circuits, it is beneficial to have custom gates only apply to
specific rows. To do this, we can add an extra column $q$ (per custom gate),
where each cell in $q$ takes the value 0 or 1. Then, we can modify the custom
gate to be
\[
q_i \cdot (a_i \cdot b_i - c_i) = 0
\]
which only applies the custom multiplication gate for rows which $q_i \neq 0$.
Column $q$ is called a selector.

\emph{Permutation arguments} allow us to constrain pairs of cells in the grid
to have equal values. They are used to copy values from one cell to another.
They are implemented via randomized polynomial constraints for \emph{multiset}
equality checks.
  
\emph{Lookup arguments} allow us to constrain a $k$-tuple of cells
$(d^1_i, \ldots, d^k_i)$ in the same row $i$ to agree with \emph{some} row of
a separate set of $k$ columns in the grid.  This constrains
$(d^1_i, \ldots, d^k_i)$ to lie in the \emph{lookup table} defined by those
$k$ other columns. We use lookup arguments in the arithmetization in two ways.
First, we implement range checks on a cell $c$ by constraining it to take
values in a fixed range $\{0, \ldots, N - 1\}$. Second, we implement
non-linearities by looking up a pair of cells $(a, b)$ in a table defined
by exhaustive evaluation of the non-linearity. Lookup arguments are also
implemented by randomized polynomial constraints. 

\vspace{0.5em}

Prior work on SNARK-ing neural networks using proof systems intended for generic
computations started with the more limited R1CS arithmetization
\cite{gennaro2013quadratic} and the Groth16 proof system \cite{groth2016size},
in which neural network inference is less efficient to express. In Section
\ref{sec:opts}, we describe how to use this more expressive Plonkish
arithmetization to efficiently express DNN inference.

\minihead{Measuring performance for Halo2}
Halo2 is an instance of a polynomial interactive oracle proof (IOP)
\cite{bensasson2016interactive} made non-interactive via the Fiat-Shamir
heuristic.  In a polynomial IOP, the ZK-SNARK is constructed from column
polynomials which interpolate the values in each column.  In Halo2, these
polynomials are fed into the \emph{inner product argument} introduced
in \cite{bowe2019recursive} to generate the final ZK-SNARK.

Several different aspects of performance matter when evaluating a ZK-SNARK
proof for a computation.  First, we wish to minimize proving time for the Prover
and verification time for the Verifier.  Second, on both sides, we wish to minimize
the proof size. Although a precise cost model for these is complex in Halo2,
all of these measures generally increase with the number of rows, columns,
custom gates, permutation arguments, and lookup arguments.

\section{Constructing ZK-SNARKs for ImageNet-Scale Models}
\label{sec:opts}

We now describe our main contribution, the implementation of a ZK-SNARK proof
for MobileNetv2 inference \cite{sandler2018mobilenetv2} in \texttt{halo2}. This
requires arithmetizing the building block operations in standard convolutional
neural networks (CNNs) in the Plonkish arithmetization.

\subsection{Arithmetization}

Standard CNNs are composed of six distinct operations: convolutions, batch
normalization, ReLUs, residual connections, fully connected layers, and softmax.
We fuse the batch normalization into the convolutions and return the logits to
avoid executing softmax. We now describe our ingredients for constraining the
remaining four operations.

\minihead{Quantization and fixed-point}
Neural network inference is typically done in floating-point arithmetic, which
is extremely expensive to emulate in the prime field of arithmetic circuits.
To avoid this overhead, we focus on DNNs quantized in \texttt{int8} and
\texttt{uint8}.  For these DNNs, weights and activations are represented as
8 bit integers, though intermediate computations may involve up to 32 bit integers.

In these quantized DNN, each weight, activation, and output is stored as a tuple
$(w_{\textrm{quant}}, z, s)$, where $w_{\textrm{quant}}$ and $z$ are $8$-bit
integer weight and zero point, and $s$ is a floating point scale factor.  $z$
and $s$ are often shared for all weights in a layer, which reduces the number of
bits necessary to represent the DNN. In this representation, the
weight $w_{\textrm{quant}}$ represents the real number weight
\[
w = (w_{\textrm{quant}} - z) \cdot s.
\]

To more efficiently arithmetize the network, we replace the floating point $s$
by a fixed point approximation $\frac{a}{b}$ for $a, b \in \mathbb{N}$ and compute
$w$ via
\[
w = ((w_{\textrm{quant}} - z) \cdot a) / b,
\]
where the intermediate arithmetic is done in standard 32-bit integer arithmetic.
Our choice of lower precision values of $a$ and $b$ results in a slight accuracy
drop but dramatic improvements in prover and verifier performance.

As an example of fixed point arithmetic after this conversion, consider adding
$y = x_1 + x_2$ with zero points and scale factors $z_y, z_1, z_2$ and $s_y, s_1, s_2$,
respectively. The floating point computation
\[
  (y - z_y) \cdot s_y = (x_1 - z_1) \cdot s_1 + (x_2 - z_2) \cdot s_2
\]
is replaced by the fixed point computation 
\[
  y \approx (x_1 - z_2) \cdot \frac{a_1}{b_1} \frac{b_y}{a_y} + (x_2 - z_2)
  \cdot \frac{a_2}{b_2} \frac{b_y}{a_y} + z_y.
\]
The addition and multiplication can be done natively in the finite field, but
the division cannot. To address this, we factor the computation of each layer
into dot products and create a custom gate to verify division. We further fuse
the division and non-linearity gates for efficiency. We describe this process
below.

\minihead{Custom gates for linear layers}
MobileNets contain three linear layers (layers with only linear operations):
convolutions, residual connections, and fully connected layers. For these linear
layers, we perform the computation per activation. To avoid expensive floating
point scaling by the scale factor and the non-linearities, we combine these
operations into a single sub-circuit.

To reduce the number of custom gates, we only use two custom gates for all
convolutions, residual connections, and fully connected layers. The first
custom gate constrains the addition of a fixed number of inputs $x^j_i$ in row $i$
via
\[
c_i = \sum_{j = 1}^N x^j_i.
\]
The second custom gate constrains a dot product \emph{of fixed size} with zero
point. For constant zero point $z$, inputs $x^j_i$, weights $w^j_i$, and output
$c_i$ in row $i$, the gate implements the polynomial constraint
\[
c_i = \sum_{j=1}^N (x^j_i - z) \cdot w^j_i
\]
for a fixed $N$.  To implement dot products of length $k < N$, we constrain
$w_{k + 1}, \ldots, w_N = 0$.  For dot products of length $k > N$, we use copy
constraints and the addition gate.

While the addition gate can be represented using the dot product gate, we use
two gates for efficiency purposes. Namely, the custom addition gate can perform
an N-element addition using half as many grid cells as the dot product gate.

\minihead{Lookup arguments for non-linearities}
Consider the result of an unscaled, flattened convolution in row $i$:
\[
c_i = \sum_j x^j_i \cdot w^j_i,
\]
where $j$ indexes over the image height, width, and channels. Performing
scale factor division and (clipped) ReLU to obtain the final activation
requires computing
\[
a_i = \text{ClipAndScale}(c_i, a; b)
:= \textrm{clip}\Big(\left\lfloor \frac{c_i \cdot a}{b} \right\rfloor, 0, 255\Big).
\]
To constrain this efficiently, we apply a lookup argument and use the same value
of $b$ across layers. To do so, we first perform the division by $b$ using a
custom gate. Since $b$ is fixed, we can use the same custom gate and lookup
argument. Let $d_i = \frac{c_i \cdot a}{b}$. We then precompute the possible
values of the input/output pairs of $(d_i, a_i)$ to form a lookup table $T =
\{(c, \text{ClipAndScale}(c)) \mid c \in \{ 0, \ldots, N \}\}$. $N$ is chosen to
cover the domain, namely the possible values of $c$. We then use a lookup
argument to enforce the constraint $\textrm{Lookup}[(d_i, a_i) \in T]$.

We emphasize that naively using lookup arguments would result in a different
lookup argument per layer, since the scale factors differ. Using different
lookup arguments would add high overhead, which our approach avoids.

\minihead{Automated translation from TensorFlow Lite}
We created a translation layer to compile TensorFlow Lite models into circuits in the
\texttt{halo2} software package.  The translation layer automatically unrolls the
inference computation into an arithmetic circuit in the Plonkish arithmetization using
the custom gate and lookup arguments described above.

Our translation layer implements two optimizations.  First, to minimize the number
of columns and number of custom gates, our translation layer avoids creating new custom
gates until there are no more available rows in existing ones. Second, we reduce the
number of lookup arguments by sharing lookup tables between layers when the scale factors
are the same. This is particularly useful for the residual layers, where the scaling factor
can be normalized to be shared across layers.

\subsection{Committing to weights or inputs}

As described in Section~\ref{sec:snarks}, ZK-SNARKs allow parts of the inputs to
be made public, in addition to revealing the outputs of the computation. For ML
models, the input (e.g., image), weights, or both can be made public.  Then, to
commit to the hidden inputs, the hash can be computed within the ZK-SNARK and be
made public.  Concretely, we use the following primitives:
\begin{enumerate}
  \item \textbf{Hidden input, public weights}: the input is hidden and the
  weights are public. The input hash is computed and made public.
  \item \textbf{Public input, hidden weights}: the input is public and the
  weights are hidden. The weight hash is computed and made public.
  \item \textbf{Hidden input, hidden weights}: the inputs and weights are
  hidden. The hash of both are computed and made public.
\end{enumerate}

To compute the hashes, we use an existing circuit for the SNARK-friendly
Poseidon hash \cite{grassi2019poseidon}. The hash of the inputs, weights, or
both can be SNARK-ed as described.

\section{Applications of Verified ML Model Inference}
\label{sec:protocols}

Building upon our efficient ZK-SNARK constructions, we now show that it is
possible to verify ML model accuracy, verify ML model predictions for serving,
and trustlessly retrieve documents matching a predicate based on an ML model.

\subsection{Protocol Properties and Security Model}

\minihead{Protocol properties}
In this section, we describe and study the properties of protocols leveraging
verified ML inference. Each protocol has a different set of requirements, which
we denote $A$. The requirements $A$ may be probabilistic (e.g., the model has
accuracy 80\% with 95\% probability). 

We are interested in the \emph{validity} and \emph{viability} of our protocols.
Validity that if the protocol completes, $A$ holds. Viability refers to the
property that rational agents will participate in the protocol.

% For the accuracy verification and inference protocols, the parties can choose to
% participate or not. As such, we are also interested in the \emph{viability} of
% our protocols, which states that economically rational actors will participate
% in our protocols.

\minihead{Security model}
In this work, we use the standard ZK-SNARK security model for the ZK-SNARKs
\cite{bunz2020transparent}. Informally, the standard security model states the
prover and verifier only interact via the ZK-SNARKs and that the adversary is
computationally bounded, which excludes the possibility of side channels. Our
security model allows for malicious adversaries, which is in contrast to the
semi-honest adversary setting. Recall that the in semi-honest adversary setting,
the adversaries honestly follow the protocol but attempt to compromise privacy,
which is common in the MPC setting.

% We further assume the standard security model of public, permissionless,
% Turing-complete smart contract execution engines (which are typically
% blockchains) \cite{mohanta2018overview}. Informally, this assumption means
% that two parties can securely
% send messages and execute smart contracts publicly. 

\minihead{Assumptions}
For validity, we only assume two standard cryptographic assumptions. First, that
it is hard to compute the order of random group elements
\cite{bunz2020transparent}, which is implied by the RSA assumption
\cite{rivest1978method}. Second, that finding hash collisions is difficult
\cite{rogaway2004cryptographic}. Only requiring cryptographic hardness
assumptions is sometimes referred to as \emph{unconditional}
\cite{ghodsi2017safety}.

For viability, we assume the existence of a programmatic escrow service and that
all parties are economically rational. In the remainder of this section, we
further assume the ``no-griefing condition,'' which states that no party will
purposefully loses money to hurt another party, and the ``no-timeout condition,"
which states that no parties will time out. Both of these conditions can be
relaxed. We describe how to relax these conditions in the
Appendix.

% For viability, we assume that all parties are economically rational
% and that there is an escrow service. We note that this excludes ``griefing''
% attacks, where a malicious party purposefully loses money to hurt another party.
% We defer the analysis of non-economically rational actors to future work. The
% escrow service can be constructed programmatically, such as via smart contracts.
% Importantly, the escrow service can also be run permissionlessly and trustlessly.

\subsection{Verifying ML model accuracy}
In this setting, a model consumer (MC) is interested in verifying a model
provider's (MP) model's accuracy, and MP desires to keep the weights hidden.
As an example use case, MC may be interested in verifying the model accuracy to
purchase the model or to use MP as an ML-as-a-service provider (i.e., to
purchase predictions in the future). Since the weights are proprietary, MP
desires to keep the weights hidden. The MC is interested in \emph{verifiable}
accuracy guarantees, to ensure that the MP is not lazy, malicious, or serving
incorrect predictions.

% For this setting, we are interested in analyzing both the validity and
% viability. As such, we analyze the pseudononymous marketplace setting, where
% there are many potential MCs and MPs. We assume the existence of an escrow
% service. This escrow service need only be able to follow programmatic rules and
% have the ability to adjudicate disputes.

Denote the cost of obtaining a test input and label to be $E$, the cost of
ZK-SNARKing a single input to be $Z$, and $P$ to be the cost of performing
inference on a single data point. We enforce that $E > Z > P$. Furthermore, let
$N = N_1 + N_2$ be the number of examples used in the verification protocol.
These parameters are marketplace-wide and are related to the security of the
protocol.

The protocol requires that MP stakes $1000 N_1 E$ per model to participate. The
stake is used to prevent Sybil attacks, in which a single party fakes the
identity of many MPs. Given the stake, the verification protocol is as follows
for some accuracy target $a$:
\begin{enumerate}
  \item MP commits to an architecture and set of weights (by providing the
  ZK-SNARK keys and weight hash respectively). MC commits to a test set $\{
  (x_1, y_1), ..., (x_N, y_N) \}$ by publishing the hash of the examples.

  \item MP and MC escrows $2 N E + \epsilon$, where $\epsilon$ goes to the
  escrow service.

  \item MC sends the test set to MP. MP can continue or abort at this point. If
  MP aborts, MC loses $N P$ of the escrow.

  \item MP sends ZK-SNARKs and the outputs of the model on the test set to MC.

  \item If accuracy target $a$ is met, MC pays $2 N Z$. Otherwise,
  MP loses the full amount $2NE$ to MC.
\end{enumerate}

The verification protocol is valid because MP must produce the outputs of the ML
model as enforced by the ZK-SNARKs. MC can compute the accuracy given the
outputs. Thus, if the protocol completes, the accuracy target is met.

If the economic value of the transaction exceeds $1000 N_1 E$, the protocol is
viable since the MP will economically benefit by serving or selling the model.
This follows as we have chosen the stake parameters so that malicious aborting
will cost the MC or MP more in expectation than completing the protocol. We
formalize our analysis and give a more detailed analysis the Appendix.

\subsection{Verifying ML Model Predictions}
In this setting, we assume that MC has verified model accuracy and is interested
in purchasing predictions in the ML-as-a-service setting. As we show, MC need
not request a ZK-SNARK for every prediction to bound malicious MP behavior.

The serving verification procedure proceeds in rounds of size $K$ (i.e.,
prediction is served over $K$ inputs). MC is allow to contest at any point
during the round, but not after the round has concluded. Furthermore, let $K
\geq K_1 > 0$. The verification procedure is as follows:
\begin{enumerate}
  \item MC escrows $2KZ$ and MP escrows $\beta KZ$, where $\beta \geq 2$ is
  decided between MP and MC.

  \item MC provides the hashes for the $K$ inputs to the escrow and sends the
  inputs to MP ($x_i$). MP verifies the hashes.

  \item MP provides the predictions ($y_i$) to the inputs (without ZK-SNARKs) to
  MC. MC provides the hash of $\textrm{Concat}(x_i, y_i)$ to the escrow.

  \item If MC believes MP is dishonest, MC can contest on any subset $K_1$ of
  the predictions.

  \item When contested, MP will provide the ZK-SNARKs for the $K_1$ predictions.
  If MP fails to provide the ZK-SNARKs, then it loses the full $\beta ZP$.

  \item If the ZK-SNARKs match the hashes, then MC loses $2 K_1 Z$ from the
  escrow and the remainder of the funds are returned. Otherwise, MP loses the
  full $\beta ZP$ to MC.
\end{enumerate}

For validity, if MP is honest, MC cannot contest successfully and the input and
weight hashes are provided. Similarly, if MC is honest and contests an invalid
prediction, MP will be unable to produce the ZK-SNARK.

For viability, first consider an honest MP. The honest MP is
indifferent to the escrow as it receives the funds back at the end of the round.
Furthermore, all contests by MC will be unsuccessful and MP gains $K_1 Z$ per
unsuccessful contest.

For honest MC to participate, they must either have a method of detecting
invalid predictions with probability $p$ or they can randomly contest a $p$
fraction of the predictions. Note that for random contests, $p$ depends on the
negative utility of MC receiving an invalid prediction. As long as $\beta KZ$ is
large relative to $\frac{K Z}{p}$, then MC will participate.

\begin{table*}[ht!]
\centering
\begin{tabular}{llllll}
Model & Accuracy (top-5) & Setup time & Proving time & Verification time & Proof
size (bytes) \\
\hline
MobileNet, 0.35, 96  & 59.1\% & 93.9s   & 163.2s  & 0.74s  & 6528 \\
MobileNet, 0.5, 224  & 75.7\% & 937.7s  & 1530.7s & 6.32s  & 7552 \\
MobileNet, 0.75, 192 & 79.2\% & 1341.2s & 2457.5s & 10.27s & 5952
\end{tabular}
\caption{Accuracy, setup time, proving time, and verification time of various
MobileNet v2 configurations. The first parameter is the ``expansion size''
parameter for the MobileNet and the second parameter is image resolution. As
shown, it is now possible to SNARK ImageNet models, which no prior work can
achieve.}
\label{table:eval}
\end{table*}

\subsection{Trustless Retrieval of Items Matching a Predicate}
In this setting, a requester is interested in retrieving records that match the
output of an ML model (i.e., a predicate) from a responder. These situations
often occur during legal subpoenas, in which a judge requires the responder to
send a set of documents matching the predicate. For example, the requester may
be a journalist requesting documents under the Freedom of Information Act or the
plaintiff requesting documents for legal discovery. This protocol could also be
useful in other settings where the responder wishes to prove that a dataset does
not contain copyrighted content.

When a judge approves this request, the responder must divulge documents or
images that match the request. We show that ZK-SNARKs allow requests encoded as
ML algorithms can be trustlessly verified.

The protocol proceeds as follows:
\begin{enumerate}
  \item The responder commits to the dataset by producing hashes of the documents.
  \label{audit:hash}
  \item The requester sends the model to the responder.
  \item The responder produces ZK-SNARKs of the model on the documents, with the
  inputs hashed. The responder sends the requester the documents that match the
  positive class of the model.
\end{enumerate}

The audit protocol guarantees the following: the responder will return the
documents from Stage \ref{audit:hash} that match the model's positive class.
The validity follows from the difficulty of finding hash collisions and the
security of ZK-SNARKs.

The responder may hash invalid documents (e.g., random or unrelated images),
which the protocol makes no guarantees over. This can be mitigated based on
whether the documents come from a trusted or untrusted source.

For documents from a trusted source, the hashes can be verified from a signature
from the trusted source. As an example, hashes for government-produced documents
(in the FOIA setting) may be produced at the time of document creation.

For documents from an untrusted source (e.g., the legal discovery setting), we
require a commitment for the entire corpus. Given the commitment, the judge can
allow the requester to randomly sample a small number ($N$) of the documents to
verify the hashes. In this case, the requester can verify that the responder
tampered with at most $p = \exp\left( \frac{1 - \delta}{N} \right)$ for some
confidence level $\delta$.

\section{Evaluation}
\label{sec:eval}

To evaluate our ZK-SNARK system, we ZK-SNARKed MobileNets with varying
configurations. We evaluated the hidden model and hidden input setting, which
is the most difficult setting for ZK-SNARKs.

We measured four metrics: model accuracy, setup time, proving
time, and verification time. The setup time is done once per MobileNet
architecture and is independent of the weights. The proving is done by the model
provider and the verification is done by the model consumer. Proving and
verification must be done once per input. To the best of our knowledge, no prior
work can ZK-SNARK DNNs on ImageNet scale models.

As mentioned, we ZK-SNARK quantized DNNs, which avoids floating point
computations. We use the model provided by TensorFlow Slim
\cite{silberman2018tf}. MobileNet v2 has two adjustable parameters: the
``expansion size'' and the input dimension. We vary these parameters to see the
effect on the ZK-SNARKing time and accuracy of the models.

\subsection{ZK-SNARKs for ImageNet-scale models}
We first present results when creating ZK-SNARKs for only the DNN execution,
which all prior work on ZK-SNARKs for DNNs do. Namely, we do not commit to the
model weights in this section. We use the AWS \texttt{r6i.32xlarge} instance
type for all experiments in this section.

\begin{table}
  \centering
  \begin{tabular}{ll}
    Method & Proving time \\
           & lower bounds (s) \\
    \hline
    Zen   & 20,000 \\
    vCNN  & 172,800 \\
    pvCNN & 31,011$^{*}$ \\ % 557.96 * 40 + 1448.83 * 6 (layers + relu lower bound)
    zkCNN & 1,597$^{*}$ % (40 / 13) * 57.7 * (96 / 32)^2 + 30.6 
  \end{tabular}
  \caption{Lower bounds on the proving time for prior work. These lower bounds
  were obtained by finding a DNN with strictly fewer operations compared to
  MobileNet v2 (0.35, 96) in the papers reporting Zen and vCNN. For pvCNN and
  zkCNN, we estimate the lower bound by scaling the computation.}
  \label{table:proving-prior-work}
\end{table}

% \minihead{End-to-end results}
We summarize results for various MobileNet v2 configurations in
Table~\ref{table:eval}. As shown, we can achieve up to 79\% accuracy on
ImageNet, while simultaneously taking as few as 10s and 5952 bytes to verify.
Furthermore, the ZK-SNARKs can be scaled down to take as few as 0.7s to verify
at 59\% accuracy. These results show the feasibility of ZK-SNARKing
ImageNet-scale models.

In contrast, we show the lower bounds on the time for prior work to ZK-SNARK a
comparable model to MobileNet v2 (0.35, 96). We were unable to reproduce any of
the prior work, but we use the proving numbers presented in the papers. For Zen,
and vCNN we use the largest model in the respective papers as lower bounds
(MNIST or CIFAR10 models). For zkCNN and pvCNN we estimate the proving time by
scaling the largest model in the paper. As shown in
Table~\ref{table:proving-prior-work}, the proving time for the prior work is at
least 10$\times$ higher than our method and up to 1,000$\times$ higher. We
emphasize that these are lower bounds on the proving time for prior work.

Finally, we note that the proof sizes of our ZK-SNARKs are orders of magnitude
less than MPC methods, which can take tens to hundreds of gigabytes.

% \minihead{Ablations}
% To see the effect of our optimizations, we removed the resource sharing between
% layers and created custom gates per convolution. We benchmarked the same model
% configurations with our optimizations removed, with results shown in
% Table~\ref{TODO}. TODO: run experiments, show results

\subsection{Protocol Evaluation}
We present results when instantiating the protocols described in
Section~\ref{sec:protocols}. To do so, we ZK-SNARK MobileNet v2 (0.35, 96)
\emph{while} committing to the weights, which \emph{no other prior work does}.
For the DNNs we consider, the cost of committing to the weights via hashes is
approximately the cost of the inference itself. This phenomena of hashing being
proportional to the computation cost also holds for other ZK-SNARK applications
\cite{zkevm}.

For each protocol, we compute the cost using public cloud hardware for the
prover and verifier for a variety of protocol parameters. We use a
cost-optimized instance for these experiments (AWS \texttt{r8i.8xlarge}). A full
deployment of ZK-SNARKs would require analyzing the assorted infrastructure
costs associated with the deployment, which is outside the scope of this work.

\begin{table}[t!]
  \centering
  \begin{tabular}{lll}
  Fraction & Sample size & Cost \\
  \hline
  5\%   & 72  & \$11.99\\
  2.5\% & 183 & \$30.48 \\
  1\%   & 366 & \$60.96
  \end{tabular}
  \caption{Costs of performing verified prediction and trustless retrieval while
  bounding the fraction of predictions tampered with. Cost were estimated with
  the MobileNet v2 (0.35, 96) model.}
  \label{table:verified-prediction-costs}
\end{table}

\minihead{Verifying prediction and trustless retrieval}
For both the verifying MP predictions and trustless retrieval, the MC
(requester) can bound the probability that the MP (responder) returns incorrect
results by sampling at random. In both cases, if a single incorrect example is
found, the MC (requester) has recourse. In the verified predictions setting, MC
will financially gain and in the retrieval setting, the requester can force the
judge to make the responder turn over all documents.

As such, the MC can choose a confidence level $\delta$ and a bound
on the fraction of predictions tampered $p$. The MC can then choose a random
sample of size $N$ as determined by inverting a valid Binomial proportion
confidence interval. Namely, \emph{$N$ is independent of the size of the
batch}.

We compute the number of samples required and the cost of the ZK-SNARKs (both
the proving and verifying) at various $p$ at $\delta = 5\%$, with results in
Table~\ref{table:verified-prediction-costs}. We use the Clopper-Pearson exact
interval \cite{clopper1934use} to compute the sample size.

To contextualize these results, consider the Google Cloud Vision API. Google
Cloud Vision charges \$1.50 per 1,000 images. Predictions over one million
images would cost \$1,500. If we could scale ZK-SNARKs to verify the Google API
model with cost on par with MobileNet v2 (0.35, 96), verifying these predictions
would add 4\% overhead, which is acceptable in many circumstances.

\begin{table}[t!]
  \centering
  \begin{tabular}{lll}
  $\epsilon$ & Sample size & Total cost \\
  \hline
  5\%   & 600    & \$99.93   \\
  2.5\% & 2,396  & \$399.08  \\
  1\%   & 14,979 & \$2494.90
  \end{tabular}
  \caption{Cost of verifying the accuracy of an ML model within some $\epsilon$
  of the desired accuracy. Costs were estimated with the MobileNet v2 (0.35, 96)
  model.}
  \label{table:verified-accuracy-costs}
\end{table}

\minihead{Verifying model accuracy}
For verifying MP model accuracy, the MC is interested in bounding probability
that the accuracy target $a$ is not met:
\[
  P(a' < a) \leq \delta
\]
for the estimated accuracy $a'$ and some confidence level $\delta$. We focus on
binary accuracy in this evaluation.

For binary accuracy, we can use Hoeffding's inequality to solve for the sample
size:
\[
  P(a - a' > \epsilon) \leq \exp\left( \frac{-2 \epsilon^2}{N} \right) = \delta
\]

We show the total number of samples needed for various $\epsilon$ at $\delta =
5\%$ and the associated costs in Table~\ref{table:verified-accuracy-costs}.
Although these costs are high, they are within the realm of possibility. For
example, it may be critical to verify the accuracy of a financial model or a
model used in healthcare settings. For reference, even moderate size datasets
can cost on the order of \$85,000 \cite{incze2019cost}, so verifying the model
would add between 0.1\% to 2.9\% overhead compared to just the cost of obtaining
training data.

\section{Conclusion}

In this work, we present protocols for verifying ML model execution trustlessly
for audits, testing ML model accuracy, and ML-as-a-service inference. We further
present the first ZK-SNARKed ImageNet-scale model to demonstrate the feasibility
of our protocols. Combined, our results show the promise for verified ML model
execution in the face of malicious adversaries.

\bibliography{paper}

\begin{thebibliography}{37}
\providecommand{\natexlab}[1]{#1}
\providecommand{\url}[1]{\texttt{#1}}
\expandafter\ifx\csname urlstyle\endcsname\relax
  \providecommand{\doi}[1]{doi: #1}\else
  \providecommand{\doi}{doi: \begingroup \urlstyle{rm}\Url}\fi

\bibitem[Armknecht et~al.(2015)Armknecht, Boyd, Carr, Gj{\o}steen, J{\"a}schke,
  Reuter, and Strand]{armknecht2015guide}
Armknecht, F., Boyd, C., Carr, C., Gj{\o}steen, K., J{\"a}schke, A., Reuter,
  C.~A., and Strand, M.
\newblock A guide to fully homomorphic encryption.
\newblock \emph{Cryptology ePrint Archive}, 2015.

\bibitem[Ben-Sasson et~al.(2016)Ben-Sasson, Chiesa, and
  Spooner]{bensasson2016interactive}
Ben-Sasson, E., Chiesa, A., and Spooner, N.
\newblock Interactive oracle proofs.
\newblock In Hirt, M. and Smith, A. (eds.), \emph{Theory of Cryptography}, pp.\
   31--60, Berlin, Heidelberg, 2016. Springer Berlin Heidelberg.
\newblock ISBN 978-3-662-53644-5.

\bibitem[Ben-Sasson et~al.(2018)Ben-Sasson, Bentov, Horesh, and
  Riabzev]{bensasson2018scalable}
Ben-Sasson, E., Bentov, I., Horesh, Y., and Riabzev, M.
\newblock Scalable, transparent, and post-quantum secure computational
  integrity.
\newblock Cryptology ePrint Archive, Paper 2018/046, 2018.
\newblock URL \url{https://eprint.iacr.org/2018/046}.
\newblock \url{https://eprint.iacr.org/2018/046}.

\bibitem[Bitansky et~al.(2017)Bitansky, Canetti, Chiesa, Goldwasser, Lin,
  Rubinstein, and Tromer]{bitansky2017hunting}
Bitansky, N., Canetti, R., Chiesa, A., Goldwasser, S., Lin, H., Rubinstein, A.,
  and Tromer, E.
\newblock The hunting of the snark.
\newblock \emph{Journal of Cryptology}, 30\penalty0 (4):\penalty0 989--1066,
  2017.

\bibitem[Bowe et~al.(2019)Bowe, Grigg, and Hopwood]{bowe2019recursive}
Bowe, S., Grigg, J., and Hopwood, D.
\newblock Recursive proof composition without a trusted setup.
\newblock Cryptology ePrint Archive, Paper 2019/1021, 2019.
\newblock URL \url{https://eprint.iacr.org/2019/1021}.
\newblock \url{https://eprint.iacr.org/2019/1021}.

\bibitem[B{\"u}nz et~al.(2020)B{\"u}nz, Fisch, and
  Szepieniec]{bunz2020transparent}
B{\"u}nz, B., Fisch, B., and Szepieniec, A.
\newblock Transparent snarks from dark compilers.
\newblock In \emph{Annual International Conference on the Theory and
  Applications of Cryptographic Techniques}, pp.\  677--706. Springer, 2020.

\bibitem[Clopper \& Pearson(1934)Clopper and Pearson]{clopper1934use}
Clopper, C.~J. and Pearson, E.~S.
\newblock The use of confidence or fiducial limits illustrated in the case of
  the binomial.
\newblock \emph{Biometrika}, 26\penalty0 (4):\penalty0 404--413, 1934.

\bibitem[Feng et~al.(2021)Feng, Qin, Zhang, Ding, and Chu]{feng2021zen}
Feng, B., Qin, L., Zhang, Z., Ding, Y., and Chu, S.
\newblock Zen: An optimizing compiler for verifiable, zero-knowledge neural
  network inferences.
\newblock \emph{Cryptology ePrint Archive}, 2021.

\bibitem[Gabizon(2021)]{gabizon2021from}
Gabizon, A.
\newblock From airs to raps - how plonk-style arithmetization works.
\newblock 2021.
\newblock URL
  \url{https://hackmd.io/@aztec-network/plonk-arithmetiization-air}.

\bibitem[Gabizon \& Williamson(2020)Gabizon and Williamson]{gabizon2020plookup}
Gabizon, A. and Williamson, Z.~J.
\newblock plookup: A simplified polynomial protocol for lookup tables.
\newblock \emph{Cryptology ePrint Archive}, 2020.

\bibitem[Gabizon et~al.(2019)Gabizon, Williamson, and
  Ciobotaru]{gabizon2019plonk}
Gabizon, A., Williamson, Z.~J., and Ciobotaru, O.
\newblock Plonk: Permutations over lagrange-bases for oecumenical
  noninteractive arguments of knowledge.
\newblock \emph{Cryptology ePrint Archive}, 2019.

\bibitem[Gennaro et~al.(2013)Gennaro, Gentry, Parno, and
  Raykova]{gennaro2013quadratic}
Gennaro, R., Gentry, C., Parno, B., and Raykova, M.
\newblock Quadratic span programs and succinct nizks without pcps.
\newblock In \emph{Annual International Conference on the Theory and
  Applications of Cryptographic Techniques}, pp.\  626--645. Springer, 2013.

\bibitem[Ghodsi et~al.(2017{\natexlab{a}})Ghodsi, Gu, and
  Garg]{ghodsi2017safety}
Ghodsi, Z., Gu, T., and Garg, S.
\newblock Safetynets: Verifiable execution of deep neural networks on an
  untrusted cloud.
\newblock 2017{\natexlab{a}}.
\newblock \doi{10.48550/ARXIV.1706.10268}.
\newblock URL \url{https://arxiv.org/abs/1706.10268}.

\bibitem[Ghodsi et~al.(2017{\natexlab{b}})Ghodsi, Gu, and
  Garg]{ghodsi2017safetynets}
Ghodsi, Z., Gu, T., and Garg, S.
\newblock Safetynets: Verifiable execution of deep neural networks on an
  untrusted cloud.
\newblock \emph{Advances in Neural Information Processing Systems}, 30,
  2017{\natexlab{b}}.

\bibitem[Grassi et~al.(2019)Grassi, Khovratovich, Rechberger, Roy, and
  Schofnegger]{grassi2019poseidon}
Grassi, L., Khovratovich, D., Rechberger, C., Roy, A., and Schofnegger, M.
\newblock Poseidon: A new hash function for zero-knowledge proof systems.
\newblock Cryptology ePrint Archive, Paper 2019/458, 2019.
\newblock URL \url{https://eprint.iacr.org/2019/458}.
\newblock \url{https://eprint.iacr.org/2019/458}.

\bibitem[Groth(2016)]{groth2016size}
Groth, J.
\newblock On the size of pairing-based non-interactive arguments.
\newblock In \emph{Annual international conference on the theory and
  applications of cryptographic techniques}, pp.\  305--326. Springer, 2016.

\bibitem[Incze(2019)]{incze2019cost}
Incze, R.
\newblock The cost of machine learning projects.
\newblock 2019.
\newblock URL
  \url{https://medium.com/cognifeed/the-cost-of-machine-learning-projects-7ca3aea03a5c}.

\bibitem[Jha et~al.(2021)Jha, Ghodsi, Garg, and Reagen]{jha2021deepreduce}
Jha, N.~K., Ghodsi, Z., Garg, S., and Reagen, B.
\newblock Deepreduce: Relu reduction for fast private inference.
\newblock In \emph{International Conference on Machine Learning}, pp.\
  4839--4849. PMLR, 2021.

\bibitem[Juvekar et~al.(2018)Juvekar, Vaikuntanathan, and
  Chandrakasan]{juvekar2018gazelle}
Juvekar, C., Vaikuntanathan, V., and Chandrakasan, A.
\newblock $\{$GAZELLE$\}$: A low latency framework for secure neural network
  inference.
\newblock In \emph{27th USENIX Security Symposium (USENIX Security 18)}, pp.\
  1651--1669, 2018.

\bibitem[Knott et~al.(2021)Knott, Venkataraman, Hannun, Sengupta, Ibrahim, and
  van~der Maaten]{knott2021crypten}
Knott, B., Venkataraman, S., Hannun, A., Sengupta, S., Ibrahim, M., and van~der
  Maaten, L.
\newblock Crypten: Secure multi-party computation meets machine learning.
\newblock \emph{Advances in Neural Information Processing Systems},
  34:\penalty0 4961--4973, 2021.

\bibitem[Kumar et~al.(2020)Kumar, Rathee, Chandran, Gupta, Rastogi, and
  Sharma]{kumar2020cryptflow}
Kumar, N., Rathee, M., Chandran, N., Gupta, D., Rastogi, A., and Sharma, R.
\newblock Cryptflow: Secure tensorflow inference.
\newblock In \emph{2020 IEEE Symposium on Security and Privacy (SP)}, pp.\
  336--353. IEEE, 2020.

\bibitem[Lam et~al.(2022)Lam, Mitzenmacher, Reddi, Wei, and
  Brooks]{lam2022tabula}
Lam, M., Mitzenmacher, M., Reddi, V.~J., Wei, G.-Y., and Brooks, D.
\newblock Tabula: Efficiently computing nonlinear activation functions for
  secure neural network inference.
\newblock \emph{arXiv preprint arXiv:2203.02833}, 2022.

\bibitem[Lee et~al.(2020)Lee, Ko, Kim, and Oh]{lee2020vcnn}
Lee, S., Ko, H., Kim, J., and Oh, H.
\newblock vcnn: Verifiable convolutional neural network based on zk-snarks.
\newblock \emph{Cryptology ePrint Archive}, 2020.

\bibitem[Liu et~al.(2021)Liu, Xie, and Zhang]{liu2021zkcnn}
Liu, T., Xie, X., and Zhang, Y.
\newblock Zkcnn: Zero knowledge proofs for convolutional neural network
  predictions and accuracy.
\newblock In \emph{Proceedings of the 2021 ACM SIGSAC Conference on Computer
  and Communications Security}, pp.\  2968--2985, 2021.

\bibitem[Lou \& Jiang(2021)Lou and Jiang]{lou2021hemet}
Lou, Q. and Jiang, L.
\newblock Hemet: A homomorphic-encryption-friendly privacy-preserving mobile
  neural network architecture.
\newblock In \emph{International conference on machine learning}, pp.\
  7102--7110. PMLR, 2021.

\bibitem[Mishra et~al.(2020)Mishra, Lehmkuhl, Srinivasan, Zheng, and
  Popa]{mishra2020delphi}
Mishra, P., Lehmkuhl, R., Srinivasan, A., Zheng, W., and Popa, R.~A.
\newblock Delphi: A cryptographic inference service for neural networks.
\newblock In \emph{29th USENIX Security Symposium (USENIX Security 20)}, pp.\
  2505--2522, 2020.

\bibitem[Mohassel \& Zhang(2017)Mohassel and Zhang]{mohassel2017secureml}
Mohassel, P. and Zhang, Y.
\newblock Secureml: A system for scalable privacy-preserving machine learning.
\newblock In \emph{2017 IEEE symposium on security and privacy (SP)}, pp.\
  19--38. IEEE, 2017.

\bibitem[Pearson et~al.(2022)Pearson, Fitzgerald, Masip, Bell{\'e}s-Mu{\~n}oz,
  and Mu{\~n}oz-Tapia]{pearson2022plonkup}
Pearson, L., Fitzgerald, J., Masip, H., Bell{\'e}s-Mu{\~n}oz, M., and
  Mu{\~n}oz-Tapia, J.~L.
\newblock Plonkup: Reconciling plonk with plookup.
\newblock \emph{Cryptology ePrint Archive}, 2022.

\bibitem[Pentyala et~al.(2021)Pentyala, Dowsley, and
  De~Cock]{pentyala2021privacy}
Pentyala, S., Dowsley, R., and De~Cock, M.
\newblock Privacy-preserving video classification with convolutional neural
  networks.
\newblock In \emph{International conference on machine learning}, pp.\
  8487--8499. PMLR, 2021.

\bibitem[Privacy \& Explorations(2022)Privacy and Explorations]{zkevm}
Privacy and Explorations, S.
\newblock zkevm, 2022.
\newblock URL
  \url{https://github.com/privacy-scaling-explorations/zkevm-circuits}.

\bibitem[Rivest et~al.(1978)Rivest, Shamir, and Adleman]{rivest1978method}
Rivest, R.~L., Shamir, A., and Adleman, L.
\newblock A method for obtaining digital signatures and public-key
  cryptosystems.
\newblock \emph{Communications of the ACM}, 21\penalty0 (2):\penalty0 120--126,
  1978.

\bibitem[Rogaway \& Shrimpton(2004)Rogaway and
  Shrimpton]{rogaway2004cryptographic}
Rogaway, P. and Shrimpton, T.
\newblock Cryptographic hash-function basics: Definitions, implications, and
  separations for preimage resistance, second-preimage resistance, and
  collision resistance.
\newblock In \emph{International workshop on fast software encryption}, pp.\
  371--388. Springer, 2004.

\bibitem[Sandler et~al.(2018)Sandler, Howard, Zhu, Zhmoginov, and
  Chen]{sandler2018mobilenetv2}
Sandler, M., Howard, A., Zhu, M., Zhmoginov, A., and Chen, L.-C.
\newblock Mobilenetv2: Inverted residuals and linear bottlenecks.
\newblock In \emph{Proceedings of the IEEE conference on computer vision and
  pattern recognition}, pp.\  4510--4520, 2018.

\bibitem[Silberman \& Guadarrama(2018)Silberman and
  Guadarrama]{silberman2018tf}
Silberman, N. and Guadarrama, S.
\newblock Tf-slim: A high level library to define complex models in tensorflow,
  2018.

\bibitem[Thaler(2013)]{thaler2013time}
Thaler, J.
\newblock Time-optimal interactive proofs for circuit evaluation.
\newblock Cryptology ePrint Archive, Paper 2013/351, 2013.
\newblock URL \url{https://eprint.iacr.org/2013/351}.
\newblock \url{https://eprint.iacr.org/2013/351}.

\bibitem[Weng et~al.(2022)Weng, Weng, Tang, Yang, Li, and Liu]{weng2022pvcnn}
Weng, J., Weng, J., Tang, G., Yang, A., Li, M., and Liu, J.-N.
\newblock pvcnn: Privacy-preserving and verifiable convolutional neural network
  testing.
\newblock \emph{arXiv preprint arXiv:2201.09186}, 2022.

\bibitem[zcash(2022)]{halo2}
zcash.
\newblock halo2, 2022.
\newblock URL \url{https://zcash.github.io/halo2/}.

\end{thebibliography}
\bibliographystyle{icml2020}

\clearpage

\appendix
\section{Viability of Verifying Model Accuracy}
\label{apx:model-acc-viability}

In this section, we prove the viability of the simplified protocol for verifying
model accuracy.

As mentioned, viability further requires that the cost of the model or price of
post-verification purchased predictions is greater than $1000 N_1 E$. Viability
requires that honest MP/MC will participate and that dishonest MP/MC will not
participate.

Consider the case of an honest MP. If MC is dishonest, it can economically gain
by having MP proceed beyond Stage \ref{verify:mp-fooled} and having MP fail the
accuracy target. However, as MP has access to the test set, they can determine
the accuracy before proceeding beyond \ref{verify:mp-fooled}, so will not
proceed if the accuracy target is not met. If MP has a valid model, they will
proceed, since the profits of serving predictions or selling the model is larger
than their stake.

Consider the case of an honest MC. Note that an economically rational MP is
incentivized to serve the model if it has a model of high quality. Thus, we
assume dishonest MPs do not have model that achieves the accuracy target. The
dishonest MP can economically gain by aborting at Stage \ref{verify:mp-fooled}
at least 1000 times (as $E > P$). MC can choose to participate with MP that only
has a failure rate of at most 1\%. In order to fool honest MCs, MP must collude
to verify invalid test sets, which costs $2 \epsilon$ per verification. MP must
have 99 fake verifications for one failed verification from an honest MC. Thus,
by setting $\epsilon = \frac{N P}{99}$, dishonest MP will not participate.

From our analysis, we see that honest MP and MC are incentivized to participate
and that dishonest MP and MC will not participate, showing viability.

\section{Verifying ML Model Accuracy with Griefing and Timeouts}
\label{apx:verified-model-acc-long}

In this section, we describe how to extend our model accuracy protocol to
account for griefing and timeouts. Griefing is when an adversarial party
purposefully performs economically disadvantageous actions to harm another
party. Timeouts are when either the MP or MC does not continue with the protocol
(whether by choice or not) without explicitly aborting.

Denote the cost of obtaining a test input and label to be $E$, the cost of
ZK-SNARKing a single input to be $Z$, and $P$ to be the cost of performing
inference on a single data point. We enforce that $E > Z > P$. Furthermore, let
$N = N_1 + N_2$ be the number of examples used in the verification protocol.
These parameters are marketplace-wide and are related to the security of the
protocol.

The marketplace requires MP to stake $1000 N_1 E$ per model to participate. The
stake is used to prevent Sybil attacks, in which a single party fakes the
identity of many MPs. Given the stake, the verification protocol is as follows
for some accuracy target $a$:
\begin{enumerate}
  \item MP commits to an architecture and set of weights (by providing the
  ZK-SNARK keys and weight hash respectively). MC commits to a test set $\{
  (x_1, y_1), ..., (x_N, y_N) \}$ by publishing the hash of the examples.

  \item MP and MC escrows $2 N E + \epsilon$, where $\epsilon$ goes to the
  escrow service.

  \item MP selects a random subset of size $N_1$ of the test set. If MC aborts
  at this point, MC loses the full amount in the escrow to MP. If MC continues,
  it sends the subset of examples to MP.

  \item MP chooses to proceed or abort. If MP aborts, MC loses $N_1 P$ of the
  escrow to MP and the remainder of the funds are returned to MC and MP.
  \label{verify:mp-fooled}

  \item MC sends the remainder of the $N_2$ examples to MP. If MP aborts from
  here on out, MP loses the full amount in the escrow ($2 N E$) to MC.

  \item MP sends SNARKs of the $N_2$ examples with outputs revealed. The weights
  and inputs are hashed.

  \item If accuracy target $a$ is met, MC pays $2 (N_1 P + N_2 Z)$. Otherwise,
  MP loses the full amount $2NE$ to MC.
\end{enumerate}

\minihead{Validity and viability (no griefing or timeouts)}
The verification protocol is valid because MP must produce the outputs of the ML
model as enforced by the ZK-SNARKs. MC can compute the accuracy given the
outputs. Thus, if the protocol completes, the accuracy target is met.

Viability further requires that the cost of the model or price of
post-verification purchased predictions is greater than $1000 N_1 E$. We must
show that honest MP/MC will participate and that dishonest MP/MC will not
participate. We first show viability without griefing or timeouts and extend our
analysis below.

Consider the case of an honest MP. If MC is dishonest, it can economically gain
by having MP proceed beyond Stage \ref{verify:mp-fooled} and having MP fail the
accuracy target. Since MP chooses the subsets $N_1$ and $N_2$, they can be drawn
uniformly from the full test set. Thus, MP can choose to proceed only if $P(a
\textrm{ met} | N_1) > 1 - \alpha$ is such that expected value for MP is
positive, where $\alpha$ depends on the choice of $\epsilon$ (we provide
concrete instantiations for $\alpha$ and $\epsilon$ below). If MC is honest, MP
gains in expected value by completing the protocol, as its expected gain is
\[
  (1 - \alpha) (N_1 P + 2 N_2 Z - \epsilon) + \alpha N_1 P.
\]

Consider the case of an honest MC. Note that an economically rational MP is
incentivized to serve the model if it has a model of high quality. Thus, we
assume dishonest MPs do not have model that achieves the accuracy target. The
dishonest MP can economically gain by aborting at Stage \ref{verify:mp-fooled}
at least 1000 times (as $E > P$). MC can choose to participate with MP that only
has a failure rate of at most 1\%. In order to fool honest MCs, MP must collude
to verify invalid test sets, which costs $2 \epsilon$ per verification. MP must
have 99 fake verifications for one failed verification from an honest MC. Thus,
by setting $\epsilon = \frac{N_1 P}{99}$, dishonest MP will not participate. For
this choice of $\epsilon$, $\alpha > \frac{49 N_1 P}{49 N_1 P + 99 NE}$.

From our analysis, we see that honest MP and MC are incentivized to participate
and that dishonest MP and MC will not participate, showing viability.

\minihead{Accounting for griefing}
We have shown that there exist choices of $\alpha$ and $\epsilon$ for viability
with economically rational actors. However, we must also account for griefing,
where an economically irrational actor harms themselves to harm another party.
It is not possible to making griefing impossible. However, we can study the
costs of griefing. By making these costs high, our protocol will discourage
griefing. In order to make these costs high, we let $\epsilon = N_1 P$.

We first consider griefing attacks against MC. For the choice of $\epsilon$,
dishonest MP must pay $99 N_1 P$ per honest MC it griefs. In particular, MC
loses $N_1 P$ per attack, so the cost of a griefing MP is 99$\times$ higher than
the cost to MC.

We now consider griefing attacks against an MP. Since MP can randomly sample, MP
can simply choose $\alpha$ appropriately to ensure the costs to a griefing MC is
high. In particular, the MP pays $2NE$ per successful attack. MP's expected gain
for executing the protocol is
\[
  (1 - \alpha) (2 N_2 Z) + \alpha N_1 P
\]
for the choice of $\epsilon$ above. Then, for
\[
  \alpha = \frac{\frac{1}{50} N E - 2 N_2 Z}{N_1 P - N_2 Z}
\]
the cost of griefing is 100$\times$ higher for griefing MC than MP. By choosing
$N_1$ and $N_2$ appropriately, MP can ensure the cost of griefing is high for
griefing MCs.

\minihead{Accounting for timeouts}
Another factor to consider is that either MC or MP can choose not to continue
the protocol without explicitly aborting. To account for this, we introduce a
sub-protocol for sending the data. Once the data is sent, if MP does not
continue after time period of time, MP is slashed.

The sub-protocol for data transfer is as follows:
\begin{enumerate}
  \item MC sends hashes of encrypted inputs to escrow and MP.
  \item MC sends encrypted inputs to MP.
  \item MP signs and publishes an acknowledgement of the receipt.
  \item MC publishes decryption key.
  \item MP contests that the decryption key is invalid or continues the
  protocol.
\end{enumerate}
If MC does not respond or aborts in Stages 1, 2, or 4, it is slashed. If MP does
not respond in Stages 3 or 5, it is slashed.

Validity follows from standard cryptographic hardness assumptions. Without the
decryption key, MP cannot access the data. With the decryption key, MP can
verify that the data was sent properly.

\end{document}